\documentclass[final,1p,times]{elsart}

\usepackage{graphicx}
\usepackage{amssymb}

\journal{Physics A}
\begin{document}

\begin{frontmatter}
\title{Similarity-Based Classification in Partially Labeled Networks}

\author{Qian-Ming Zhang$^{1}$,}
\author{Ming-Sheng Shang$^{1}$,}
\author{Linyuan L\"u$^{2}$}
\corauth{Email address: linyuan.lue@unifr.ch (Linyuan L\"u)}

\address{$^{1}$Web Sciences Center, School of Computer Science and Engineering,University of Electronic Science and Technology of China, 610054 Chengdu, P. R. China \\ $^{2}$Department of Physics, University of Fribourg, Chemin du Mus\'{e}e 3, CH-1700 Fribourg, Switzerland}

\begin{abstract}
We propose a similarity-based method, using the similarity between
nodes, to address the problem of classification in partially labeled
networks. The basic assumption is that two nodes are more likely to
be categorized into the same class if they are more similar. In this
paper, we introduce ten similarity indices, including five local
ones and five global ones. Empirical results on the co-purchase
network of political books show that the similarity-based method can
give high accurate classification even when the labeled nodes are
sparse which is one of the difficulties in classification.
Furthermore, we find that when the target network has many labeled
nodes, the local indices can perform as good as those global indices
do, while when the data is spares the global indices perform better.
Besides, the similarity-based method can to some extent overcome the
unconsistency problem which is another difficulty in classification.
\end{abstract}

\end{frontmatter}

\section{Introduction}
Recently, the problem of within-network classification in partial
labeled networks has attracted much attention. Given a network with
partial nodes being labeled, the problem is to predict the labels of
these unlabeled nodes based on the known labels and the network
structure. Many algorithms have been proposed. These methods can be
widely applied to many fileds, such as the hypertext categorization
\cite{Chakrabarti1998,Yang2002}, distinguishing the fraud and legit
users in cell phone network \cite{Gallagher2008}, detecting whether
an email is for a certain task \cite{Carvalho2005} and predicting
the disease-related genes \cite{Zhang2009}. Generally speaking, the
known methods can be classified into two groups. One is collective
classification, which refers to the combined classification by using
three types of correlations: (1) between the node's label and its
attributes, (ii) between node's label and its neighbor's attributes,
(iii) between node's label and its neighbor's label (see a brief
introduction in Ref. \cite{Sen2008}). One remarkably advantage of
this method is its high ability to learn the dependency structure,
such as positive or negative correlation (i.e. consistency or
unconsistency). However, when the labeled nodes are sparse, this
method is difficult to give accurate classification. The sparse
problem can be solved by another group of methods, named
semi-supervised learning, which make use of both labeled and
unlabeled data for training (see Ref. \cite{Zhu2009} for more
information). The latent assumption of this method is the
consistency with the label information, namely the nearby nodes tend
to have the same label. Therefore when this assumption does not hold
the performance of this method will be largely degraded. Brian
\emph{et al.} proposed a method by adding ghost edges between every
pair of labeled and unlabeled node to the target network, which
enable the flow of information from the labeled nodes to the
unlabeled nodes \cite{Gallagher2008}. They assigned a weight to each
ghost edge based on the score of the two endpoints obtained by the
\emph{Even-step random walk with restart} (Even-step RWR)algorithm.
The experimental results on real-world data showed that their method
can to some extent solve the sparse problem and negative correlation
problem (i.e. unconsistency), and perform well while the existing
approaches, such as collective classification and semi-supervised
learning, will fail. In this paper, we compare the performances of
Even-step RWR index with other nine similarity indices which have
been widely used in link prediction problem
\cite{Kleinberg2007,Linyuan2009,Linyuan20092}. These include five
local indices, namely the \emph{Common Neighbors}
\cite{Lorrain1971}, \emph{Jaccard coefficient} \cite{Jaccard1901},
\emph{S{\o}rensen index} \cite{Sorensen1948}, \emph{Adamic-Adar
index} \cite{Adamic2003} and \emph{Resource Allocation index}
\cite{Linyuan2009}, and four global indices, namely \emph{Katz
index} \cite{katz1953}, \emph{Average Commute Time}
\cite{Klein1993}, \emph{cosine based on the Pseudoinverse of the
Laplacian matrix} ($cos^+$) and  \emph{Random walk With Restart}
(RWR) \cite{Brin1998}. In addition, we also consider a simple
relational neighbors algorithm, which claims that an unlabeled node
tends to have the same label with its neighbors
\cite{Macskassy2003}. Empirical results on the co-purchase network
of political books show that the similarity-based methods perform
better than the relational neighbors algorithm. Especially when the
labeled nodes are sparse, the improvement is prominent. Furthermore,
when the data is dense, the local indices perform as good as the
global indices, while when the data is spare the global indices will
perform better.

The rest of this paper is organized as follows. In section 2 we
introduce ten similarity indices, including five indices based on
local information and others based on global information. Section 3
describes the metric to evaluate the algorithm's accuracy. Section 4
shows the experimental results of the ten indices on the co-purchase
network of political books. Finally, we conclude this paper in
section 5.

\section{Similarity indices}
We consider five local similarity indices as well as five global
ones. All are defined based on the network structure. A short
introduction of each index is shown as:

(1) \emph{Common Neighbors} --- For a node $x$, let $\Gamma(x)$
denote the set of neighbors of $x$. By common sense, two nodes, $x$
and $y$, are more similar if they have many common neighbors. The
simplest measure of this neighborhood overlap is the directed count,
namely
\begin{equation}
s^{CN}_{xy}=|\Gamma(x)\cap \Gamma(y)|.
\end{equation}
where $|Q|$ is the cardinality of the set $Q$. It is obvious that
$s_{xy}=(A^2)_{xy}$, where $A$ is the adjacency matrix, in which
$A_{xy}=1$ if $x$ and $y$ are directly connected and $A_{xy}=0$
otherwise. Note that, $(A^2)_{xy}$ is also the number of different
paths with length 2 connecting $x$ and $y$.

(2) \emph{Jaccard Index} \cite{Jaccard1901} --- This index was
proposed by Jaccard over a hundred years ago, and is defined as
\begin{equation}
s^{Jaccard}_{xy}=\frac{|\Gamma(x)\cap \Gamma(y)|}{|\Gamma(x)\cup
\Gamma(y)|}.
\end{equation}

(3) \emph{S{\o}rensen Index} \cite{Sorensen1948} --- This index is
used mainly for ecological community data, and is defined as
\begin{equation}
s^{S{\o}rensen}_{xy}=\frac{2\times |\Gamma(x)\cap
\Gamma(y)|}{k(x)+k(y)}.
\end{equation}

(4) \emph{Adamic-Adar Index} \cite{Adamic2003} --- This index
refines the simple counting of common neighbors by assigning the
less-connected neighbors more weight, and is defined as:
\begin{equation}
s^{AA}_{xy}=\sum_{z\in \Gamma(x)\cap \Gamma(y)}\frac{1}{\texttt{log}
k(z)}.
\end{equation}

(5) \emph{Resource Allocation} \cite{Linyuan2009}--- Consider a pair
of nodes, $x$ and $y$, which are not directly connected. The node
$x$ can send some resource to $y$, with their common neighbors
playing the role of transmitters. In the simplest case, we assume
that each transmitter has a unit of resource, and will equally
distribute it between all its neighbors. The similarity between $x$
and $y$ can be defined as the amount of resource $y$ received from
$x$, which is:
\begin{equation}
s^{RA}_{xy}=\sum_{z\in \Gamma(x)\cap \Gamma(y)}\frac{1}{k(z)}.
\end{equation}
Clearly, this measure is symmetric, namely $s_{xy}=s_{yx}$. Note
that, although resulting from different motivations, the AA index
and RA index have the very similar form. Indeed, they both depress
the contribution of the high-degree common neighbors in different
ways. AA index takes the $logk(z)$ form while RA index takes the
linear form. The difference is insignificant when the degree, $k$,
is small, while it is great when $k$ is large. Therefor, RA index
punishes the high-degree common neighbors heavily.

(6) \emph{Katz Index} \cite{katz1953} --- This measure is based on
the ensemble of all paths, which directly sums over the collection
of paths and exponentially damped by length to give the short paths
more weights. The mathematical expression reads
\begin{equation}
s^{Katz}_{xy}=\sum^{\infty}_{l=1}\beta^l\cdot|paths^{<l>}_{xy}|=\beta{A}+\beta^2A^2+\beta^3A^3+\cdots,
\end{equation}
where $paths^{<l>}_{xy}$ is the set of all paths with length $l$
connecting $x$ and $y$, and $\beta$ is a free parameter controlling
the weights of the paths. Obviously, a very small $\beta$ yields a
measure close to CN, because the long paths contribute very little.
The $S$ matrix can be written as $(I-\beta A)^{-1}-I$. Note that,
$\beta$ must be lower than the reciprocal of the maximum of the
eigenvalues of matrix $A$ to ensure the convergence.

(7) \emph{Average Commute Time} \cite{Klein1993} --- Denoting by
$m(x,y)$ the average number of steps required by a random walker
starting form node $x$ to reach node $y$, the average commute time
between $x$ and $y$ is $n(x,y)=m(x,y)+m(y,x)$, which can be computed
in terms of the Pseudoinverse of the Laplacian matrix $L^{+}$, as:
\begin{equation}
n(x,y)=E(l_{xx}^{+}+l_{yy}^{+}-2l_{xy}^{+}),
\end{equation}
where $l_{xy}^{+}$ denotes the corresponding entry in $L^{+}$.
Assuming two nodes are considered to be more similar if they have a
small average commute time, then the similarity between the nodes
$x$ and $y$ can be defined as the reciprocal of $n(x,y)$, namely
\begin{equation}
s^{ACT}_{xy}=\frac{1}{l_{xx}^{+}+l_{yy}^{+}-2l_{xy}^{+}}.
\end{equation}

(8) \emph{Cosine based on $L^{+}$} \cite{Klein1993} --- This index
is an inner-product based measure, which is defined as the
\textbf{cosine} of node vectors, namely
\begin{equation}
s_{xy}^{cos^{+}}=cos(x,y)^{+}=\frac{l_{xy}^{+}}{\sqrt{l_{xx}^{+}\cdot{l_{yy}^{+}}}}.
\end{equation}

(9) \emph{Random walk with restart} \cite{Brin1998} --- This index
is a direct application of the PageRank algorithm. Consider a random
walker starting from node $x$, who will iteratively moves to a
random neighbor with probability $c$ and return to node $x$ with
probability $1-c$. Denote by $q_{xy}$ the probability this random
walker locates at node $y$ in the steady state, then we have
\begin{equation}\label{rwr}
\vec{q_x}= c P^{T} \vec{q_x}+(1-c)\vec{e_x},
\end{equation}
where $\vec{e_x}$ is an $N\times 1$ vector with the $x^{th}$ element
equal to $1$ and others all equal to $0$, and $P^T=AD^{-1}$ where
$D_{ij}=\delta_{ij}k_i$. The solution is straightforward, as
\begin{equation}
\vec{q_x} = (1-c) (I-cP^{T})^{-1}\vec{e_x}.
\end{equation}
Then we define the similarity between node $x$ and node $y$ equals
$s_{xy}=q_{xy}+q_{yx}$.

(10) \emph{Even-step RWR} \cite{Gallagher2008} --- To avoid the
immediate neighbors, we can consider only the even-length paths.
Mathematically, we should replace the transition matrix with
$M=(P^T)^2$.

For comparison, we compare the above-mentioned ten indices with the
simplest method, says Relational Neighbors (RN)
\cite{Macskassy2003}. Given an unlabeled node $u$, the probability
that its label is $l_i$ equals
\begin{equation}
p(l_i|u)=\frac{|V'|_{\{v'_i\in\Gamma(u)|label(v'_i)=l_i\}}}{|V''|_{\{v''_i\in\Gamma(u)|label(v''_i)\neq{\varnothing}\}}},
\end{equation}
where $V'$ is the set constituted by $u$'s neighbors whose label is
$l_i$, and $V''$ is the set of $u$'s neighbors being labeled.

\begin{figure}
\begin{center}
\includegraphics[width=13cm]{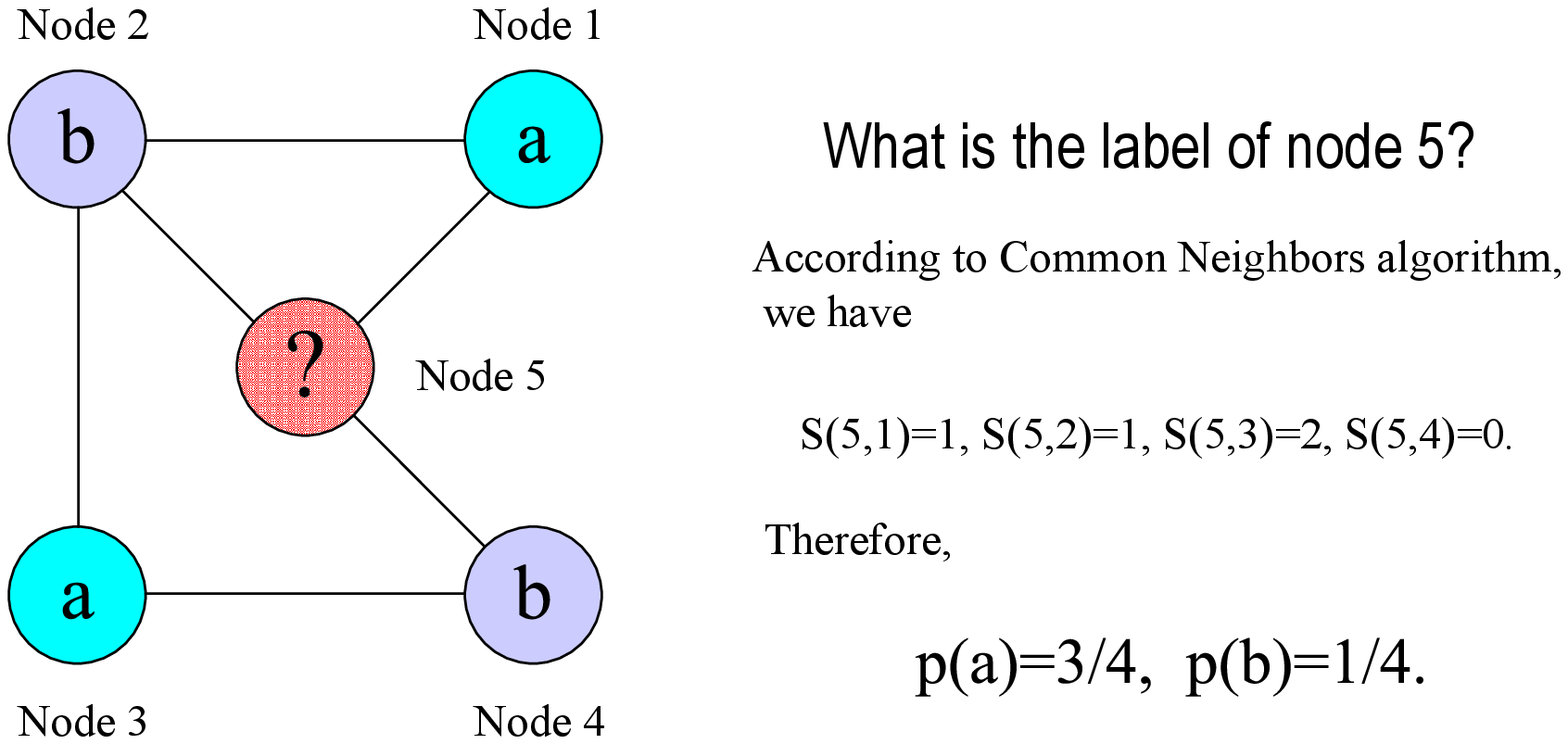} \caption{(Color
online) An illustration of how to predict the node's label according
to the similarity.}\label{1}
\end{center}
\end{figure}

\section{Method}
Consider an unweighted undirected network of both labeled and
unlabeled nodes: $G(V, E, L)$, where $V$ is the set of nodes, $E$ is
the set of links and $L=\{l_1,l_2,\cdots,l_m\}$ is the set of
labels. For each pair of nodes, $x$ and $y$, every algorithm
referred in this paper assigns a score as $s_{xy}$. For an unlabeled
node $u$, the probability that it belongs to $l_i$ is
\begin{equation}
p(l_i|u)=\frac{\sum_{\{v|label(v)=l_i\}}s_{u,v}}{\sum_{\{v|label(v)\neq{\varnothing}\}}s_{u,v}},
\end{equation}
where $l_i\in{L}$. The predicted label of node $u$ is determined by
the largest $p(l_i|u)$. If there are more than one maximum values,
we randomly select one. A simple example is shown in Fig.~\ref{1},
where there are two kinds of labels (i.e. $a$ and $b$) and five
nodes, four of which are labeled already. Our task is to predict the
label of the node 5. According to the common neighbors algorithm, we
obtain the similarity between node 5 and the other four labeled
nodes, and then we infer that the probability that node 5 is labeled
by $a$ equals $3/4$.

To test the algorithm's accuracy, all the labeled nodes are randomly
divided into two parts: the training set, $V^T$, is treated as known
information, while the probe set, $V^P$, is used for testing. We
denote $q$ the proportion of labeled nodes divided into training
set, which is considered as the density index. A smaller $q$
indicates a sparser labeled network. The accuracy is quantified by
the probability that we predict right. For a testing node
$u\in{V^P}$ whose label is $l_i$, if $p(l_i)>p(l_j), j\neq{i}$, we
predict right, and thus $q_u=1$. If there is $n$ maximum values
corresponding to $n$ different labels and the right label is one of
them, we have $q_u=1/n$. Run over all the testing nodes we have the
accuracy equals
\begin{equation}
Accuracy=\frac{\sum_{u\in{V^P}}q_u}{|V^P|},
\end{equation}
where $|V^P|$ is the number of nodes in the probe set. For example,
if there are two categories in the target network, namely $l_1$ and
$l_2$, accuracy can be obtained by
\begin{equation}
Accuracy=\frac{n'+0.5n''}{|V^P|},
\end{equation}
where $n'$ is the number of nodes in probe set being predicted right
and $n''$ is the number of nodes $u\in{V^P}$ having the same
probability of two labels (i.e. $p(l_1|u)=p(l_2|u)$).

\section{Empirical results}
We compare the above-mentioned ten similarity indices on the
co-purchases network of political books \cite{political}. This
network contains 105 nodes (books) and 441 edges. All books are
classified into three categories, neutral, liberal and conservative.
For simplicity, we start the experiments with the sampled networks
containing only two classes. Therefore, we sample three labeled
networks with three tasks as follows:

\textbf{Task 1:} \emph{Whether an unlabel node is neutral?} For this
task, we label the books which are neutral by $a$ and others by $b$
(i.e. not neutral).

\textbf{Task 2:} \emph{Whether an unlabel node is liberal?} For this
task, we label the books which are liberal by $a$ and others by $b$
(i.e. not liberal).

\textbf{Task 3:} \emph{Whether an unlabel node is conservative?} We
label the books which are conservative by $a$ and others by $b$
(i.e. not conservative).

\begin{figure}
\begin{center}
\includegraphics[width=14cm]{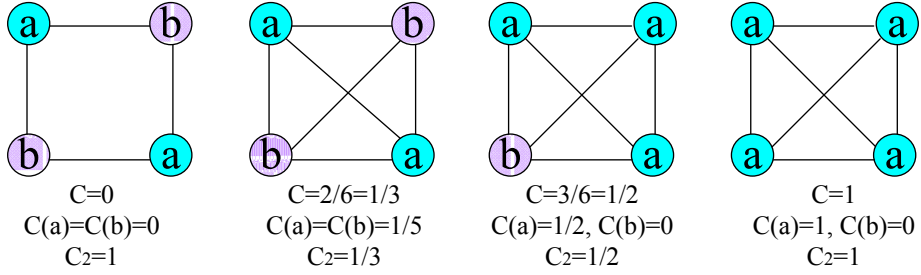}
\caption{(Color online) Illustration of the calculation of local
consistency and two-step consistency.}\label{LC}
\end{center}
\end{figure}

Table.1 summarize the basic statistics of these three sampled
networks corresponding to task 1, task 2 and task 3 respectively.
$N(x)$ ($x=a,b$) is the number of nodes labeled by $x$. $E(x)$
indicates the number of edges connecting to the nodes labeled by
$x$. Denote by $M(x)$ the number of edges whose two endpoints have
the same label $x$, then $C(x)=M(x)/E(x)$ indicats the local
consistency of the subgraph constituted by the nodes labeled by $x$
and the edges connecting to these nodes. $C$ is the local
consistency of the whole network, which reads
$C=\frac{M(a)+M(b)}{E}$, where $E$ is the total number of edges of
the whole network (here $E=441$). Note that, $E<E(a)+E(b)$. Here, we
further develop the definition of local consistency to two-step
consistency denoting by $C_2$ which equals to the number of path
with length 2 whose two endpoints have the same label divide by the
number of the path with length 2. Clearly, the common neighbor index
will perform well in the network with high $C_2$. Four simple
examples of calculating $C(x)$, $C$ and $C_2$ are shown in
Fig.~\ref{LC}. One can see that in the first graph, because of
$C=0$, RN will perform very bad, while CN performs very good
($C_2=1$). However in the forth graph both RN and CN can give good
performance.

\begin{figure}
\begin{center}
\includegraphics[width=6.8cm]{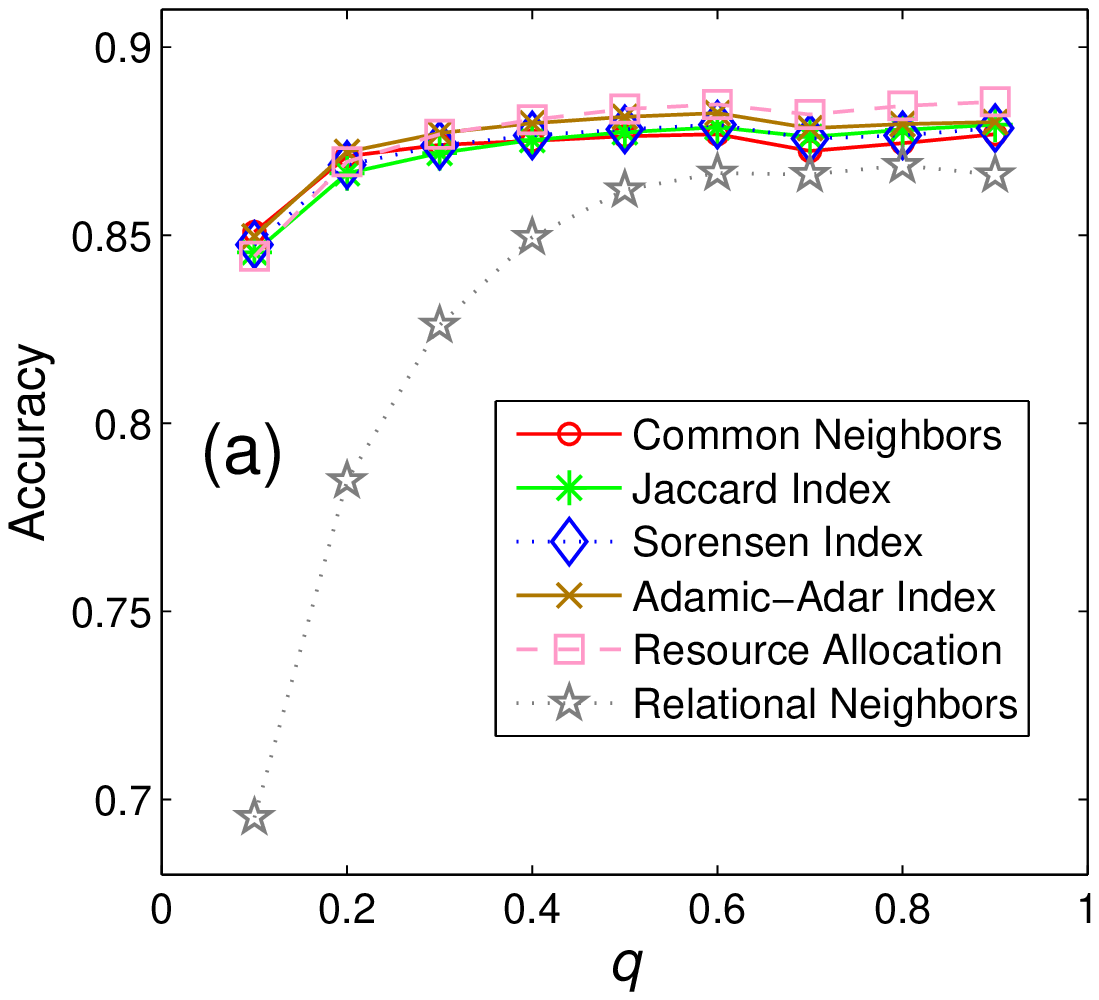}
\includegraphics[width=6.8cm]{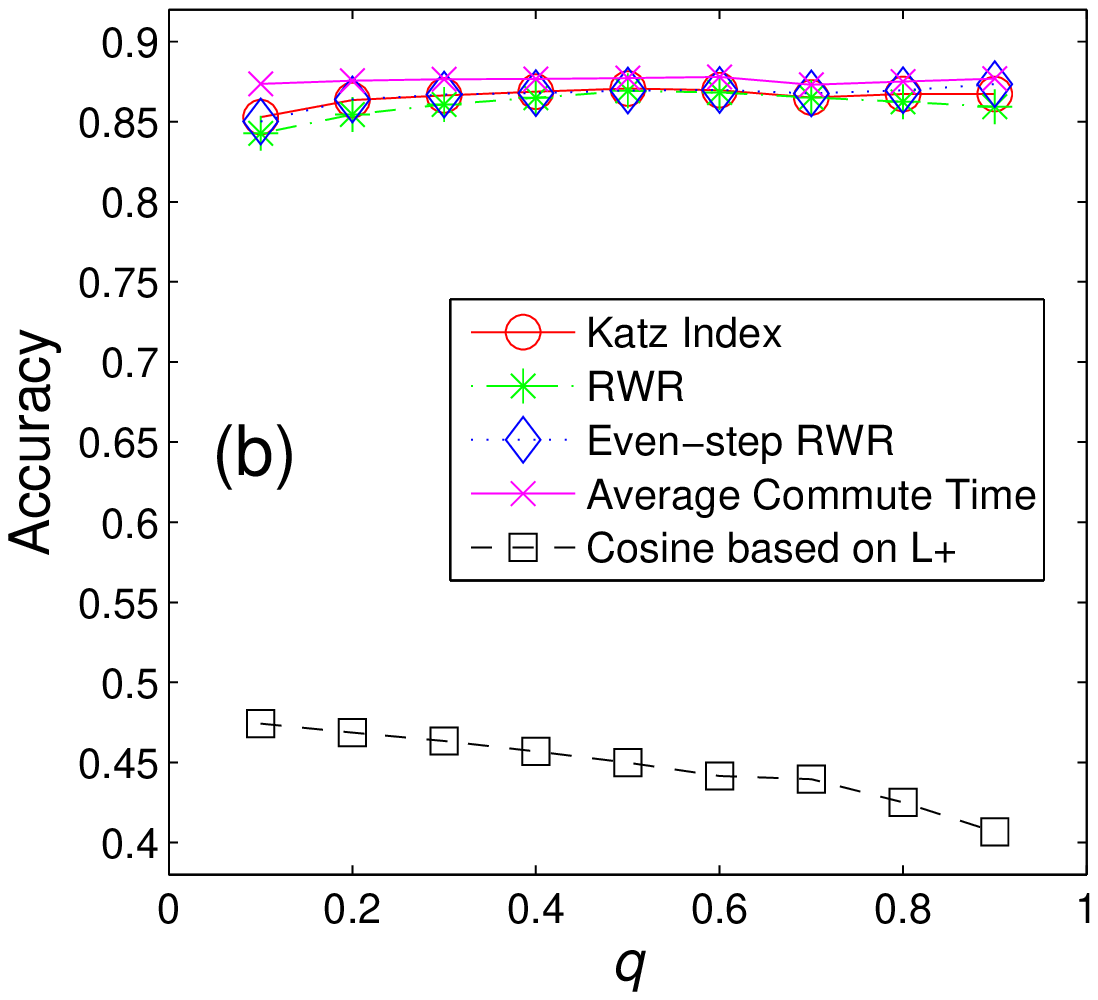}
\includegraphics[width=6.8cm]{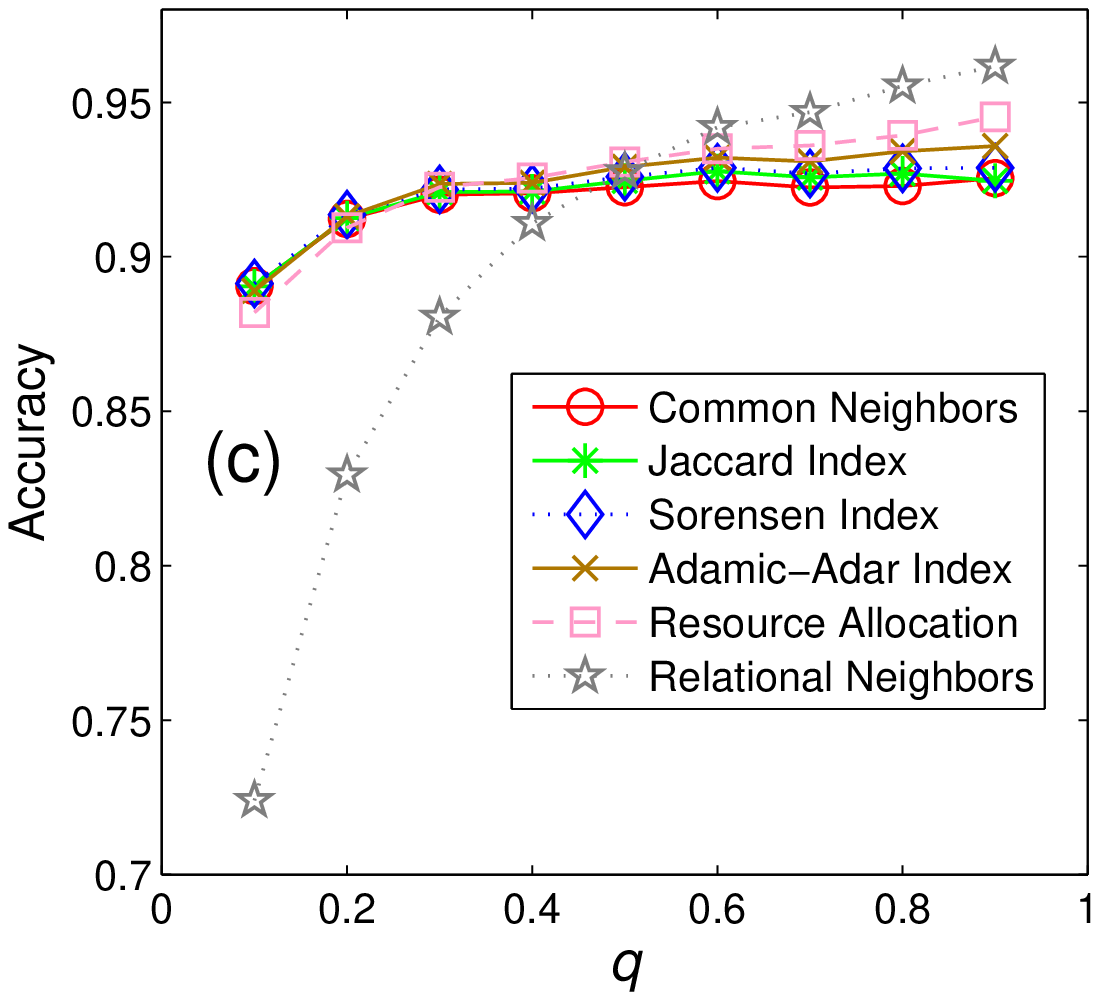}
\includegraphics[width=6.8cm]{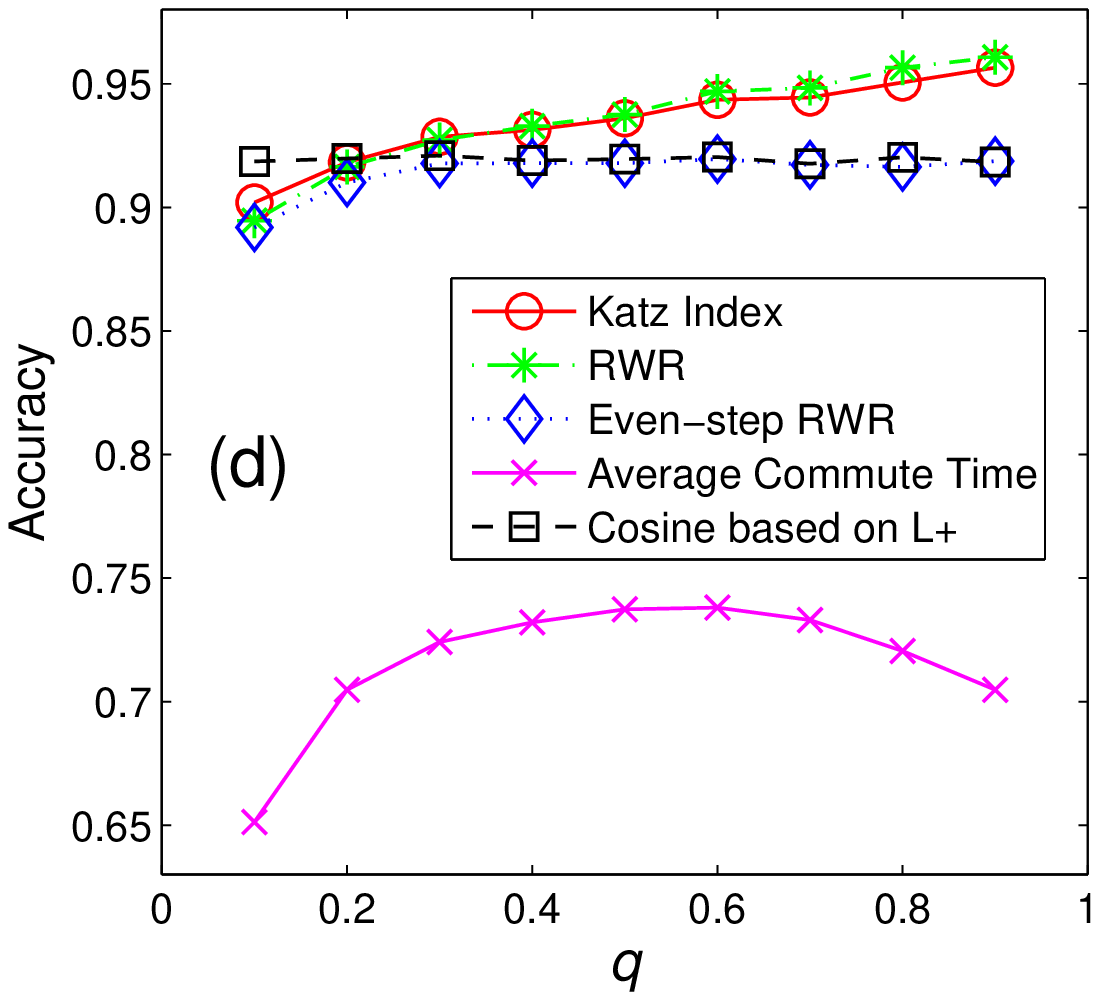}
\includegraphics[width=6.8cm]{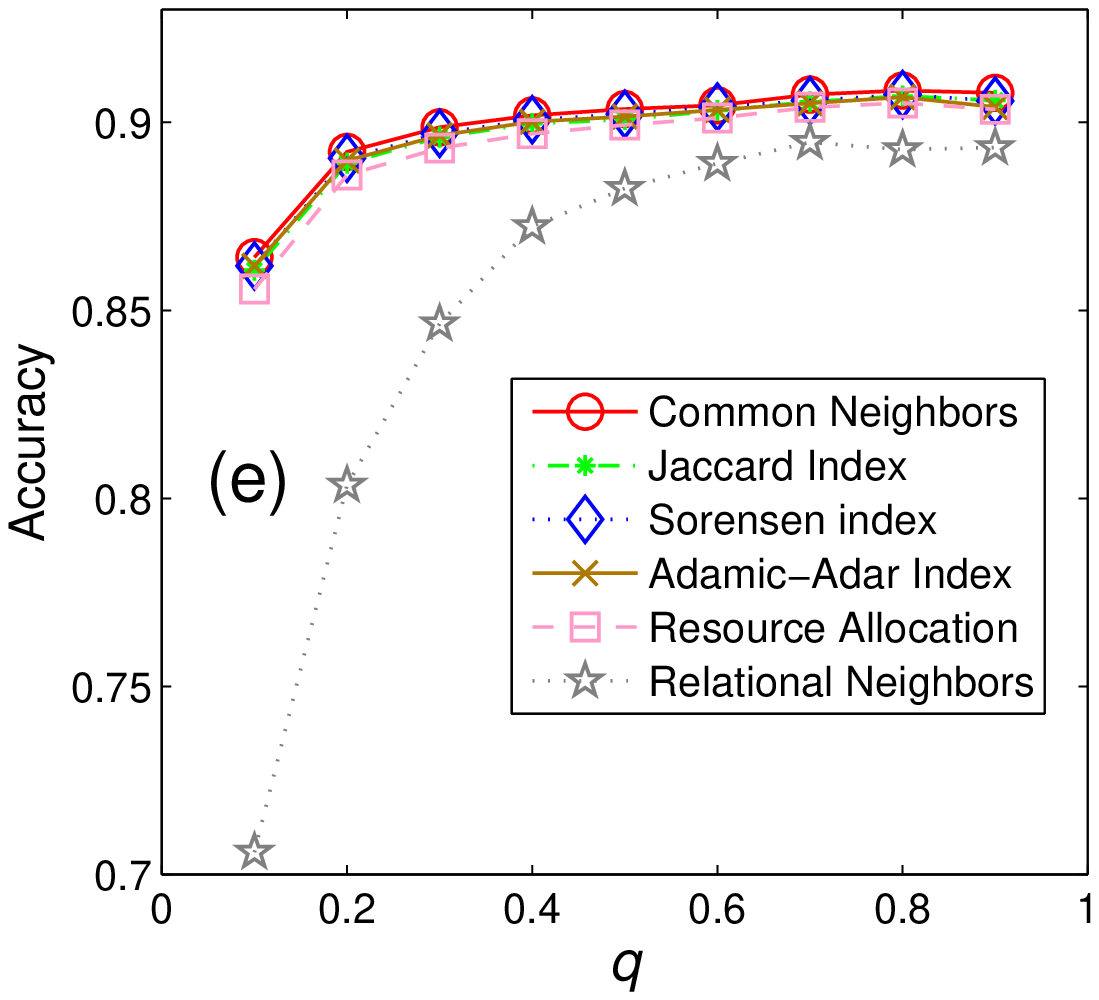}
\includegraphics[width=6.8cm]{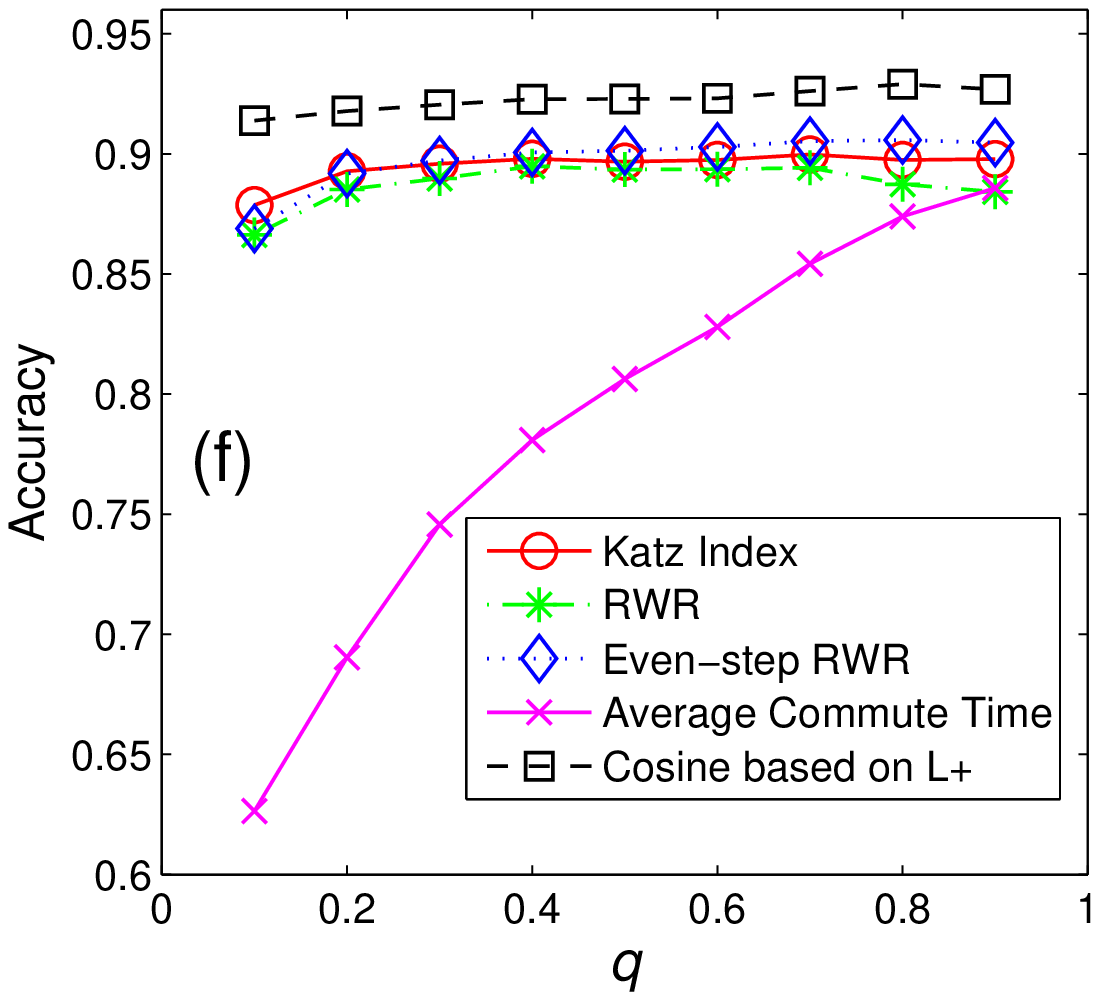}
\caption{(Color online) Comparison of ten similarity indices on
three sampled networks containing two categories. (a) and (b) are
the results of the local and global indices for task 1 respectively.
(c) and (d) are the results of the local and global indices for task
2 respectively. (e) and (f) are the results of the local and global
indices for task 3 respectively. For RWR index we set $c=0.1$. Each
number is obtained by averaging over 1000 implementations with
independently random division of training set and probe
set.}\label{sample}
\end{center}
\end{figure}

\begin{table}
\caption{The summary of local consistency of each label and each
sampled networks. $N(a)$ and $N(b)$ are the number of nodes labeled
by $a$ and $b$ respectively. $E(a)$ and $E(b)$ indicate the number
of edges connecting to the nodes labeled by $a$ and $b$
respectively. $C(a)$ and $C(b)$ are the local consistency of the
nodes labeled by $a$ and $b$ respectively. $C$ and $C_2$ are the
local consistency and two-step consistency of the sampled network,
respectively.}
\begin{center}
\begin{tabular} {ccccccccccc}
  \hline
  \hline
   Net    & $N(a)$ & $N(b)$ & $E(a)$ & $E(b)$ & $M(a)$ & $M(b)$ & $C(a)$ & $C(b)$ & $C$ & $C_2$ \\
   \hline
   Net1  & 13 & 92 & 67  & 432 & 9   & 374 & 0.134 & 0.866 & 0.869 & 0.864\\
   Net2  & 43 & 62 & 208 & 269 & 172 & 233 & 0.827 & 0.866 & 0.918 & 0.894\\
   Net3  & 49 & 56 & 236 & 251 & 190 & 205 & 0.805 & 0.817 & 0.890 & 0.882\\
   \hline
   \hline
    \end{tabular}
\end{center}
\end{table}

Comparison of the ten similarity indices on three sampled networks
are shown in Fig.~\ref{sample}. The subgraphs (a), (c) and (e) show
the results of the local indices, while (b), (d) and (f) report the
results of the global indices. It is interesting that all these five
local indices give almost the same results especially when the
density of labeled nodes is small. This is because all these five
indices are common-neighbor based and when $q$ is small whether an
unlabeled node relevant with a labeled node play a more important
role than the exact correlation (similarity score) between them.
Furthermore, because of the high $C_2$ of these three networks, all
the common-neighbor-based indices performs well and even when the
data is sparse they can give much better prediction than RN. Compare
with global indices, the local indices can give competitively
accurate classification when $q$ is large, but when the labeled data
is sparse, for most unlabeled node it is too difficult to find a
labeled node nearby, and thus the global indices will perform
better. Among these five global indices, the performance of Katz
index, RWR and even-step RWR are stable, while the performance of
ACT and $cos^+$ are not. For example, in sampled network 1, the ACT
index performs very well but $cos^+$ is even worse than pure chance.
However, in sampled network 3, the $cos^+$ index preforms the best
but the ACT index performs even worse than the simplest method RN.

\begin{figure}
\begin{center}
\includegraphics[width=6.8cm]{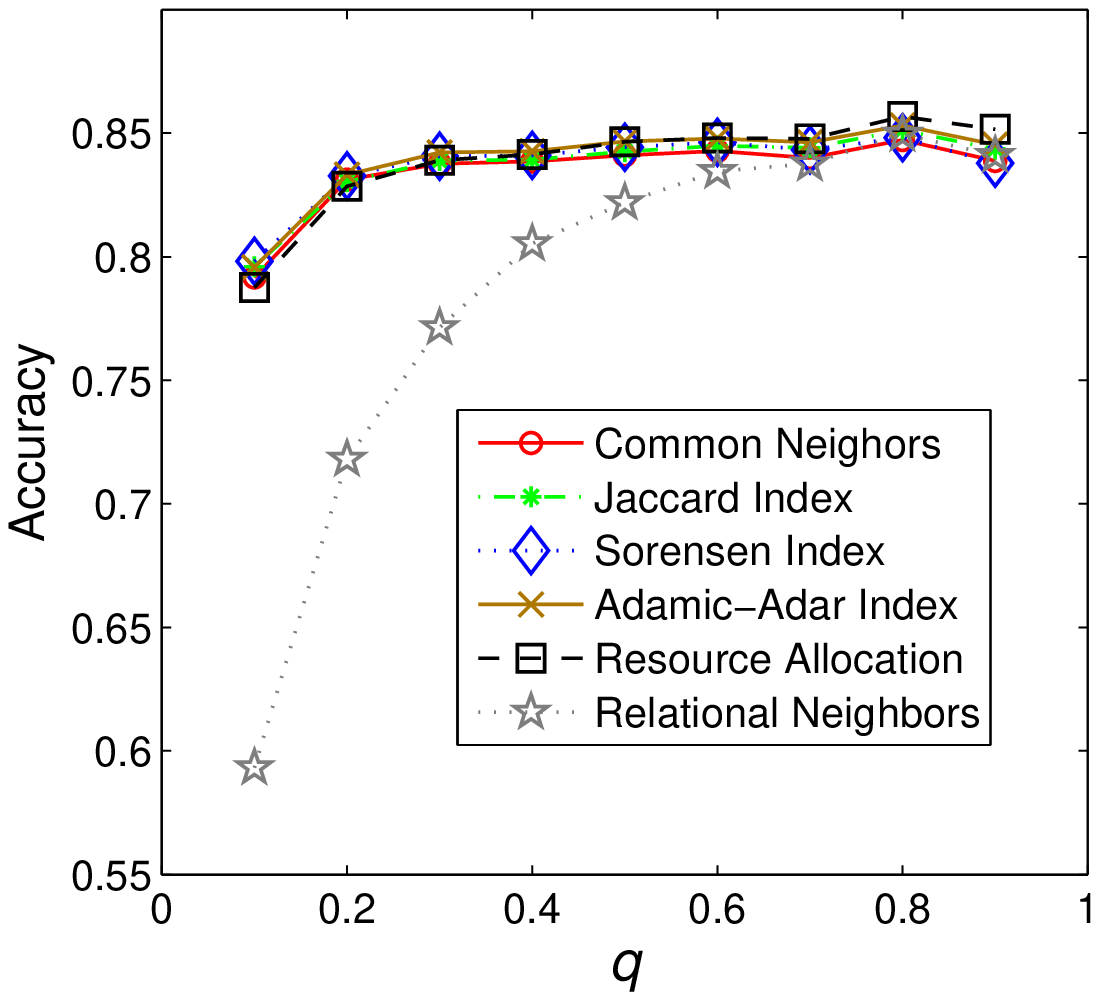}
\includegraphics[width=6.8cm]{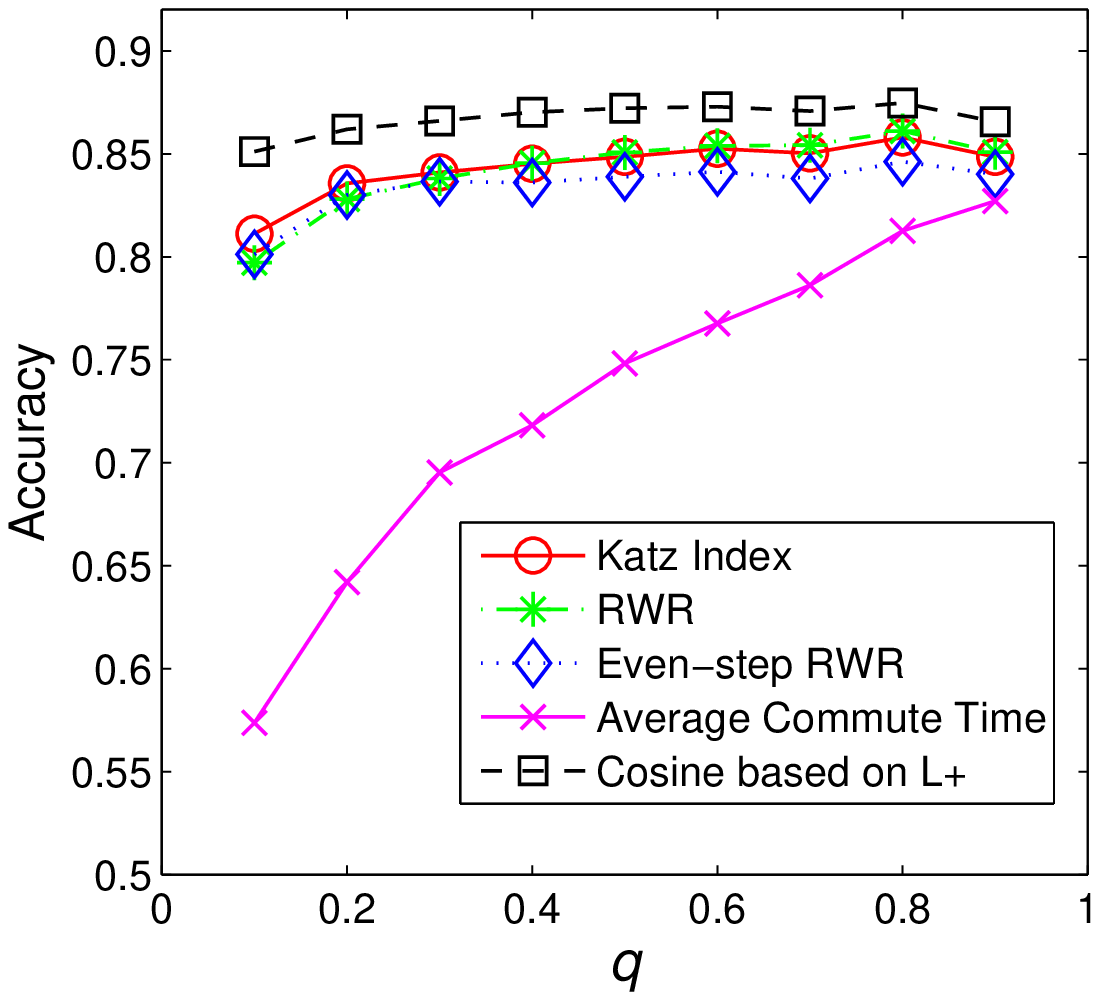}
\caption{(Color online) Comparison of ten similarity indices on the
network taking into account three categories. For RWR we set
$c=0.1$. Each number is obtained by averaging over 1000
implementations with independently random division of training set
and probe set.}\label{all}
\end{center}
\end{figure}

Obviously, it will be more difficult to obtain highly accurate
classification when we consider many categories together. We futher
carry out an experiment on the network containing all the three
categories. Our task is to detect the category of an unlabel book,
namely is it neutral, liberal or conservative? We label the books by
$n$ (i.e. neutral), $l$ (i.e. liberal) and $c$ (i.e. conservative)
according to their categories. The local consistency and two-step
consistency of this network are 0.8413 and 0.8204 respectively,
which are all lower than the three sampled networks containing only
two classes, and thus the accuracy is also lower, as shown in
Fig.~\ref{all}. One can see that the results are similar to the one
on the sampled network 3 where the biggest class, conservative, is
considered. This result demonstrates that the majorities play the
main role.

\section{Conclusion and Discussion}
In this paper, we investigated the similarity-based classification
for partial labeled network. The basic assumption is that two nodes
are more likely to have the same label if they are more similar to
each other. We introduced ten similarity indices which have been
widely used to solve the link prediction problem of complex
networks, including five common-neighbor-based indices, namely
\emph{Common Neighbors}, \emph{Jaccard coefficient},
\emph{S{\o}rensen index}, \emph{Adamic-Adar index} and
\emph{Resource Allocation index}, and five global indices, namely
\emph{Katz index}, \emph{Average Commute Time}, \emph{cosine based
on the Pseudoinverse of the Laplacian matrix} ($cos^+$),
\emph{Random walk With Restart} (RWR) and \emph{Even-step RWR}. We
carried out the experiments on the co-purchase network of political
books. The results showed that the similarity-based classification
perform much better than the relational neighbors algorithm,
especially when the labeled nodes are sparse. Furthermore, we found
that when the data is dense the local indices can perform as good as
the global indices. However, when the data is sparse, for an
unlabeled node it is too difficult to find a labeled node nearby,
and thus the global indices perform better. Compare with the former
proposed algorithms the group of similarity-based classification
methods has three advantages: firstly, it can to some extent solve
the sparse data problem by using the global indices; secondly, when
the network consistency assumption is not hold it can still give
high accurate classification; thirdly, without any learning process
this method has lower calculation complexity than other complicated
methods.

However, there are still some open problems left. For example what
is the relation between the network label structure and the
performance of each similarity index. In-depth analysis on the
modeled networks may be helpful, where we can control the label
density, network consistency and also the proportion of each class.
Anyway, we hope this work can provide a novel view for the study of
classification in partial labeled networks and we believe that there
is still a large space for further contribution. For example, when
the number of nodes in one class is much lager than in the others,
the unlabeled nodes are more likely to have the same labels with the
majority. To solve this problem we can only consider the top-$k$
similar labeled nodes when calculate the probability. In addition,
we can also use negative correlation in the adjacent matrix $A$
directly, namely for the nonzero element in $A$ if the node $x$ and
$y$ have the different labels, we set $A_{xy}=-1$. To do this, we
can not only obtain the strength of the correlation between the
unlabeled node and the labeled one but also know the correlation
type, positive or negative.

\section{ACKNOWLEDGEMENT}
We acknowledge Tao Zhou for his assistance of manuscript
preparation. This work is partially supported by the Swiss National
Science Foundation (200020-121848), the China Postdoctoral Science
Foundation (20080431273) and the National Natural Science Foundation
of China (60973069, 90924011).


\end{document}